\documentclass[twocolumn,prl]{revtex4}
\usepackage{graphicx}
\usepackage{amssymb,amsmath,bm}
\newcommand{\be}{\begin{equation}}
\newcommand{\ee}{\end{equation}}
\newcommand{\bea}{\begin{eqnarray}}
\newcommand{\eea}{\end{eqnarray}}

\newcommand{\p}{\partial}

\newcommand{\lp}{\left(}
\newcommand{\rp}{\right)}

\renewcommand{\phi}{\varphi}
\renewcommand{\epsilon}{\varepsilon}

\renewcommand{\Im}{\mathop{\rm Im}\nolimits}
\newcommand{\sgn}{\mathop{\rm sign}\nolimits}

\begin{document}
\title{Vacuum Polarization and Screening of Supercritical Impurities in Graphene}
\author{
A. V. Shytov,${}^1$ M. I. Katsnelson,${}^2$ L. S. Levitov${}^3$}
\affiliation{
${}^1$ Condensed Matter Physics and Materials Science Department, 
Brookhaven National Laboratory, Upton, 
New York 11973-5000\\
${}^2$ Radboud University of Nijmegen, Toernooiveld 1
6525 ED Nijmegen,
The Netherlands, \\
${}^3$ Department of Physics,
 Massachusetts Institute of Technology, 77 Massachusetts Ave,
 Cambridge, MA 02139
}


\begin{abstract}
Screening of charge impurities in graphene 
is analyzed using the exact solution for vacuum polarization
obtained from the massless Dirac-Kepler problem. 
For the impurity charge below certain critical value 
no density perturbation is found
away from the impurity, in agreement with the linear response theory result.
For supercritical charge, however, the polarization distribution
is shown to have a power law profile, 
leading to screening of the excess charge at large distances.
The Dirac-Kepler scattering states give rise to 
standing wave oscillations in the local density of states
which appear and become prominent in the supercritical regime.
\end{abstract}
\maketitle

Massless Dirac excitations in graphene\,\cite{reviewGK}
provide an interesting realization of quantum electrodynamics (QED) 
in dimension 
two\,\cite{Gonzalez94}. Because of zero mass and strong interactions,
characterized by a large ``fine structure constant'' 
$\alpha = e^2 /\hbar v_F\approx 2.5$,
where $v_F \approx 10^6$ m/s is the Fermi velocity, 
this material breaks away from the perturbative
QED paradigm.
One of the phenomena fundamental 
in QED, expected to become ultra-strong in graphene,
is ``vacuum polarization'' induced by charge impurities.
Scattering on charge impurities features prominently in transport properties 
of graphene, where 
it is believed to be the leading factor 
limiting electron mobility\,\cite{nomura,ando,dassarma},
providing an explanation for the conductivity\,\cite{Novoselov05}
linear in the carrier concentration. 
Although the problem of Coulomb scattering by charge
impurities received a lot of attention\,\cite{nomura,mele,ando,Katsnelson06,mirlin,dassarma}, the key question of screening
of the impurity potential outside the weak coupling regime
has not been adequately addressed\,\cite{Katsnelson06,mirlin}.

Here we present an accurate
nonperturbative treatment of this problem
based on the vacuum polarization
found from the exact solution of the 2d Dirac-Kepler problem. 
There are two qualitatively different regimes emerging from this solution,
which are somewhat analogous to those known 
in QED of heavy and superheavy atoms\,\cite{Zeldovich}.
The Coulomb potential of subcritical strength
can be treated as a mathematical singularity in solving
the Dirac equation, while in the supercritical case
a consistent solution is only possible after 
finite radius of charge distribution in a nucleus is accounted for\,\cite{Pomeranchuk}.
We shall see that a similar phenomenon takes place in our problem
at the critical charge value
\be\label{eq:beta}
\beta=\beta_c=\frac12
,\quad
\beta\equiv\frac{Ze^2}{\kappa\hbar v_F}
,
\ee
where $\kappa$ is the effective dielectric constant. 
For the case when screening 
is solely due to the graphene electrons, 
the RPA approach\,\cite{Gonzalez94} gives $\kappa_{\rm RPA}\approx 5$. 
With $e^2/\hbar v_F\approx 2.5$, this yields a critical value $Z_c\approx 1$.

The most prominent effect in our problem, arising at supercritical $\beta$,
is the change in the character of polarization of the Dirac vacuum.
While at $\beta<\frac12$ the polarization charge $q_p$ 
is localized on the scale of
the impurity radius, exhibiting no long range tail\,\cite{mirlin}, 
for supercritical $\beta$ the solution of the massless Dirac equation predicts
a power law for the spatial profile of polarization.
For $\frac12<\beta<\frac32$, when just the lowest angular momentum channels
of the Dirac equation are overcritical, we find
\be\label{eq:n_pol}
n_{\rm pol}(\rho)\approx - \frac{N\gamma\sgn\beta}{2\pi^2\rho^2}
- q_p\delta(\rho)
,\quad
\gamma\equiv \sqrt{\beta^2-{\textstyle \frac14}}
,
\ee
where $N=4$ is the combined spin and valley degeneracy of graphene.
The result (\ref{eq:n_pol}), valid for noninteracting fermions,
is somewhat modified at higher $\beta$ (see Eq.(\ref{eq:many_channels})). 


The result (\ref{eq:n_pol}) can be used to
describe screening in an interacting system in a selfconsistent renormalization group (RG) fashion.
The RG flow for polarization cloud is constructed by 
proceeding from the lattice scale $\rho=r_0$ to larger $\rho$, 
treating the net polarization charge within
radius $\rho$ as an effective point charge $\beta(\rho)$, and using it
to determine polarization at larger distances.
As a result, the net charge $\beta(\rho)$ flows from its initial value
$\beta(r_0)$ to lower values at larger distances. 
The net polarization charge (\ref{eq:n_pol}) within 
the annulus $\rho_1 < \rho < \rho_2$
equals $\delta Z = -N\sgn\beta \frac{ \gamma}{\pi} \ln (\rho_2/\rho_1)$,
which leads to the RG equation
%
\be\label{eq:RG}
\frac{d\beta(\rho)}{d\ln\rho} = 
     -\frac{N e^2\sgn\beta}{\pi\kappa \hbar v_F}\gamma(\rho)
,\quad
\beta>\beta_c
.
\ee
%
Integrating 
the flow (\ref{eq:RG}) we find that it terminates at 
a distance 
$\rho_\ast=r_0\exp\lp\frac{\pi\kappa \hbar v_F}{N e^2}\cosh^{-1}(2\beta)\rp$
where $\beta$
reaches the critical value (\ref{eq:beta}).
In contrast to screening in metals,
here the polarization 
build-up brings the net charge down to the critical value $\beta_c$ that 
remains unscreened at larger distances $\rho\gtrsim \rho_\ast$.
The RG treatment is applicable when the RG flow is slow, 
i.e., when the right-hand side of Eq.~(\ref{eq:RG}) is small. 
Thus the RG framework is adequate 
near the criticality,~$\beta \approx \beta_c$, 
where~$\gamma$ is small, 
even in the case of strong coupling, 
$e^2/\kappa \hbar v_F \sim 1$,
and the predicted termination of screening at large $\rho$ is universal.

Our treatment of vacuum polarization relies on the 
exact solution of the Dirac-Kepler problem from which we extract
scattering phases and
use them 
in the Friedel sum rule framework to evaluate the screening charge. 
The phases
are found to behave differently for $\beta<\beta_c$ and $\beta>\beta_c$.
In the first case, the essential part of the phase is
$\beta\ln k\rho$, while in the second case it is 
$\beta\ln k\rho - \gamma\sgn\beta\ln k r_0$.
It is important to realize that
the term $\beta\ln k\rho$ is the same for all
angular momentum channels. Such a contribution to the phase, as 
Ref.\cite{LL-3} insightfully remarks, arises from quasiclassical dynamics 
at large distances, and thus
has nothing to do with scattering. In agreement with this, we find that
it does not contribute to polarization at finite $\rho$, while
the term $- \gamma\sgn\beta\ln k r_0$ gives rise to the power law
in (\ref{eq:n_pol}).







Now we turn to the analysis of the massless Dirac equation 
in dimension two 
in a central potential $V(\rho)$: 
%
\begin{equation}
\hbar v_F
\left(
\begin{array}{cc}
0 & -i\partial_x - \partial_y \\
-i \partial_x + \partial_y & 0 
\end{array}
\right) 
\psi = (\epsilon - V(\rho)) \psi
. 
\label{dirac}
\end{equation}
%
Introducing polar coordinates~$x + i y = \rho e^{i\varphi}$, Ñ
we separate angular harmonics of the two-component wave function $\psi$,
and seek the solution in the form\,\footnote{
We use the ansatz $\psi\propto \rho^{s-\frac12}$, as appropriate 
for the 2d case, instead of $\psi\propto \rho^{s-1}$ 
which is a convention in 3d.}
\begin{equation}
\label{ansatz}
\psi (\rho, \varphi)  = 
\left(
\begin{array}{c}
w(\rho) + v(\rho) \\ %
\left(w(\rho) - v(\rho)\right) e^{i\varphi}
\end{array}
\right) \rho^{s - \frac12} e^{i{m}\varphi} e^{i k \rho}
,
\end{equation}
%
with integer angular quantum number ${m}$.
The terms $w$ and $v$ represent the incoming and outgoing waves.
The parameters $s$ and $k$
are determined by the behavior at small and large
$\rho$. For $V(\rho) = Ze^2/\rho$, we find
%
\begin{equation}\label{eq:k,s}
s = \sqrt{\left({m} + {\textstyle \frac12}\right)^2 - \beta^2}
,\quad 
k = -\frac{\epsilon}{\hbar v_F}, 
\end{equation}
where $\beta$ is the dimensionless coupling (\ref{eq:beta}). 
(The minus sign in (\ref{eq:k,s}) is chosen
to make $k>0$ in the Fermi sea.)
The solution (\ref{ansatz}) behaves differently 
for $|\beta| < |{m} + \frac12|$, when the exponent $s$ is real,
and $|\beta| > |{m} + \frac12|$, when $s$ becomes complex imaginary. 

The ansatz~(\ref{ansatz}), substituted into Dirac equation, yields
coupled equations for the functions~$w(\rho)$
and~$v(\rho)$:
\begin{eqnarray}\label{eq:w,u}
&& \rho \p_\rho w + \left(s  + i \beta + 2 i k\rho \right) w 
- \lp {m} + {\textstyle \frac12}\rp v = 0
, \\
\label{eq:v'}
&& \rho \p_\rho v + \left(s  - i \beta\right) v 
- \lp {m} + {\textstyle \frac12}\rp  w = 0
. 
\end{eqnarray}
We eliminate $w$ and,
after introducing a new independent variable~$z = - 2 i k \rho$, 
obtain a hypergeometric equation
\begin{equation}
\label{hyper-eq}
z v'' + (2s+1 - z) v'
  - \left(s  - i \beta \right) v = 0
. 
\end{equation}
The solution regular at~$z = 0$ is given by the confluent hypergeometric
function\,\cite{AbramowitzStegun}:
\begin{eqnarray}
\label{v-hyper}
&& v(z) = A\, {}_1F_{1} \left(s - i \beta, 2s + 1, z\right),
\\
\label{w-hyper}
&& w(z) = A\, \frac{s  - i\beta}{{m} + \frac{1}{2}}\,
  {}_1F_{1} \left(s + 1 - i \beta, 2s + 1, z\right)
,
\end{eqnarray}
where $A$ is a normalization factor.
The expression for $w$ was obtained
using Eq.(\ref{eq:v'}) and an identity for ${}_1F_1$ \cite{identity}.

The solution (\ref{ansatz}),(\ref{v-hyper}),(\ref{w-hyper}) of the Dirac-Kepler problem, regular at $\rho=0$, 
can be used to evaluate
the polarization charge in the subcritical case
$|\beta|<1/2$. 
This can be done most easily using the scattering phases,
given by the behavior of $w$ and $v$
at large $\rho$. Using the asymptotic form of the functions
${}_1F_{1}$\,\cite{asymptotic}, we find
\be
\label{wv-asymptotics}
v(\rho) = \frac{\lambda e^{ i \beta \ln (2 k \rho)}}{(2k\rho)^s}
,\quad
w(\rho)  =
\frac{\lambda^\ast e^{ -i \beta \ln (2 k \rho)} e^{-2ik\rho}}{(2k\rho)^s}
,
\ee
where the parameter $\lambda$ depends on ${m}$ and $\beta$
but not on $k$. From (\ref{wv-asymptotics}) we see that 
$\rho^{s-\frac12} w(\rho) \exp(ik\rho)$
and $\rho^{s-\frac12}v(\rho) \exp(i k \rho)$ 
indeed describe the incoming and the outgoing waves, characterized
by relative phase 
\begin{equation}
\label{plane-wave-phase}
v/w=e^{2i\delta_{m}(k) +2ik \rho}
,\quad
\delta_{m}(k) =  \beta \ln (2 k \rho) + \arg \lambda
.
\end{equation}
The log dependence in Eq.~(\ref{plane-wave-phase}) is typical for
the phase coming from~$1/r$ Coulomb tail~\cite{LL-3}. 

The scattering phases $\delta_{m}(k)$ can be used to find the
polarization charge pulled on the origin. 
However, a straightforward application
of the Friedel
sum rule\,\cite{Friedel,Lin},
involving phases evaluated at the Fermi level, 
encounters a difficulty due to the
position and energy dependence of the phases in Eq.~(\ref{plane-wave-phase}).
Since this may indicate that the polarization 
is distributed rather than localized at $\rho=0$,
we proceed with caution.

To evaluate the excess particle number $Q_{\rm pol}(\rho)$ 
in the interval $0\le \rho'<\rho$,
we note that the states with wavelength greater than $\rho$,
i.e. $|k|\lesssim 1/\rho$,  contribute negligibly to
$Q_{\rm pol}(\rho)$. Thus we can write the sum rule\,\cite{Friedel, Lin} as
\be\label{eq:Friedel_sum_rule}
Q_{\rm pol}(\rho)\approx -\frac{N}{\pi}\sum_{m} \delta_{m}(k\sim 1/\rho)
,
\ee
where the minus sign corresponds to that in Eq.(\ref{eq:k,s}).
Conveniently,
the expressions  (\ref{plane-wave-phase}) for $\delta_{m}(k)$, valid at
$k\rho\gtrsim 1$, can be used to evaluate (\ref{eq:Friedel_sum_rule}). 
However, since $\delta_{m}(k)$ depend on the product $k\rho$, 
they yield a $\rho$-independent result for $Q_{\rm pol}(\rho)$.
We therefore conclude that the polarization charge 
is concentrated on the lattice scale $\rho\approx r_0$.


\begin{figure}
\includegraphics[width=2.3in]{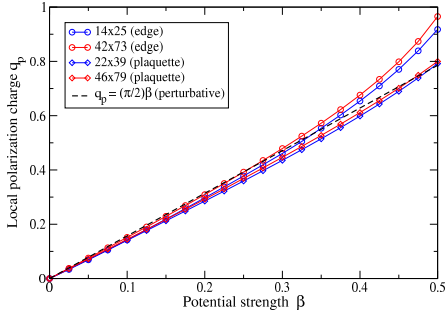} 
\vspace{-0.1cm}
\caption[]{
Local polarization charge found from numerical solution of a tight-binding 
problem on a honeycomb lattice. The charge was placed in the middle
of a rectangular region of size $n_1\times n_2$ at the lattice plaquette 
center or edge center (see legend). 
Shown is the net polarization at a distance 
of less than 5 lattice constants from point charge,
which agrees 
with the prediction
of a perturbative RPA calculation (dashed line).
}
\label{fig:polarization}
\end{figure}


To independently verify the 
conclusion about polarization at $\beta<1/2$ concentrated 
at $\rho\lesssim r_0$
we evaluated it directly using 
the eigenstates given by Eqs.~(\ref{ansatz}),(\ref{v-hyper}),(\ref{w-hyper}).
This calculation involves an energy cutoff 
introduced at the bottom of graphene band, corresponding to
$k\approx r_0^{-1}$.
We found nonvanishing contribution to polarization charge 
only on the cutoff scale,
leading to an expression $n_{\rm pol}(\rho)=-q_p\delta(\rho)$
(the second term in Eq.(\ref{eq:n_pol})).
This form of polarization charge can be independently justified by the RPA
method\,\cite{ando,mirlin}, giving $q_p=\frac{\pi}2 \beta$. Our numerical
results at $0<|\beta|\lesssim 1$, presented in Fig.\ref{fig:polarization},
yield a very similar dependence, nearly linear at $|\beta|<\frac12$,
independently confirming the above analysis.

The behavior of the scattering phase and
of the polarization charge changes when the potential strength $|\beta|$
exceeds $|{m} + \frac12|$ for one or several values of ${m}$.
For such supercritical $\beta$
Eq.(\ref{v-hyper}) is not the only possible 
solution. Adding another solution of the 
equation~(\ref{hyper-eq}), we write the function~$v(z)$
in the form
\begin{eqnarray}\label{v-fall}
v(z) &=& A\,  {}_1F_{1} \left(i (\gamma - \beta), 1 + 2 i \gamma, z\right)
 \\ \nonumber
  &+& B \,  z^{-2i\gamma} 
  {}_1F_{1}\left(-i (\gamma + \beta), 1 - 2i \gamma, z\right)
, 
\end{eqnarray}
where~$\gamma = 
\Im s  = \sqrt{\beta^2 - \left({m} + \frac12\right)^2}$, 
and~$z = -2ik\rho$. 
With the help of the relation \cite{identity} we find 
%
%
\begin{eqnarray}\label{eq:w(z)_general}
& w(z)&= -i A \eta
	  {}_1F_{1} (1 + i \gamma - i\beta, 1 + 2i \gamma, z)
\\ \nonumber
	&& -i(B /\eta)
	  z^{-2i\gamma} 
	  {}_1F_{1} (1 - i \gamma - i \beta, 1 - 2i \gamma, z)
, 
\end{eqnarray}
%
where 
$\eta= \sqrt{\frac{\beta - \gamma}{\beta + \gamma}}$. 

To find the relation between $A$ and $B$ we consider our solution
at small distances, $\rho \approx r_0\ll 1/k$:
\begin{eqnarray} 
v (\rho) &\approx& A + B e^{-\pi \gamma} e^{-2i\gamma \ln (2k\rho)}
\ , 
\\
w(\rho)  &\approx& -i A\eta
             -i (B/\eta) e^{-\pi \gamma} e^{-2i\gamma \ln (2k\rho)}     
\ .
\end{eqnarray}
Without loss of generality we use the boundary condition
$\psi_2(\rho = r_0)=0$, which enforces zero wavefunction
on one of the graphene lattice sites. (Similar boundary condition
was used to describe graphene zigzag edge 
in the Dirac equation framework\,\cite{SSC}.)
%
Solving the equation $v(r_0)=w(r_0)$, we find the relation
%
\begin{equation}
\label{eq:chi}
B = e^{i\chi(k)}\eta  e^{\pi\gamma} A
,\quad
e^{i\chi(k)}= -i \frac{1 - i\eta}{1 + i\eta} e^{2i\gamma \ln 2k r_0 }
.
\end{equation}
The product $k r_0$ is very small for typical $k$, making the phase factor
$e^{i\chi(k)}$ a rapidly oscillating function of $k$.

To better understand the role of the second solution, let us take another look
at the subcritical case, $|\beta|<|{m}+\frac12|$,
when the parameter $s$, Eq.(\ref{eq:k,s}), is real.
In this case, two independent solutions are still provided by
Eqs.(\ref{v-fall}),(\ref{eq:w(z)_general}), whereby
$i\gamma$ is replaced by $s$. 
Applying the boundary conditions the same way as above, 
instead of (\ref{eq:chi}) we find $B/A\propto (2k r_0)^{2s}\ll 1$,
which indicates that the second solution plays no role in the subcritical case.

To link 
$\chi(k)$, Eq.(\ref{eq:chi}), to the scattering phase,
we write our solution for $v(\rho)$, $w(\rho)$
at large distances $\rho k \gg 1$, again
using the asymptotic expression for ${}_1F_1$\,\cite{asymptotic},
%
%
%
\begin{equation}
\label{eq:u/w}
\frac{v}{w} = 
\frac{g_{\beta,\gamma}
+ e^{i\chi} e^{-\pi\gamma} \eta g_{\beta,-\gamma}
}{
   e^{-\pi\gamma}\eta g^\ast_{\beta,-\gamma}
 + e^{i\chi}g^\ast_{\beta,\gamma}
}
\, 
e^{2ik\rho} e^{ 2i \beta \ln 2k\rho}
,
\end{equation}
where $g_{\beta,\gamma}=\Gamma (1 + 2i \gamma)/\Gamma (1 + i \gamma + i\beta)$.
We note that (\ref{eq:u/w}) automatically
satisfies the current conservation requirement  
$|v|=|w|$. 
The relative phase of $v$ and $w$, 
defined as in Eq.(\ref{plane-wave-phase}), thus equals
%
\begin{equation}
\label{eq:xi_caseII}
\delta_{m}(k) =\theta(k) + \beta \ln 2 k\rho + \arg g_{\beta,\gamma} 
,
\end{equation}
where 
%
\begin{equation}
\label{chi(k)}
\theta(k) = 
\arg \left[e^{-\frac{i}2\chi(k)} + a e^{\frac{i}2\chi(k)}\right]
,\quad
a= e^{-\pi\gamma}
\eta \frac{g_{\beta,-\gamma}}{g_{\beta,\gamma}}
.
\end{equation}
The last two terms of the phase (\ref{eq:xi_caseII}) are
identical in form to  (\ref{plane-wave-phase}). 
They represent spherical wave ``deformed'' 
by the tail of Coulomb potential at large distances
and, as we discussed above, 
give no contribution to polarization charge at finite $\rho$.
%
The term $\theta(k)$, however, arising from the boundary
condition at small~$\rho$ via the phase~$\chi = \arg(B/A)$, Eq.(\ref{eq:chi}), 
makes the behavior completely different from that found for 
$|\beta|<|{m}+\frac12|$.

The phase $\theta(k)$ dependence on  $k$, arising via the
phase $\chi(k)$, is quite peculiar. 
To understand the relation $\theta$ {\it vs.} $\chi$,
we find the winding number of $\theta(k)$ that depends on
%
\be
|a|=
\sqrt{\frac{e^{-2\pi\gamma}-e^{-2\pi\beta}}{e^{2\pi\gamma}-e^{-2\pi\beta}}}
=
\begin{cases} <1 & \text{if $\beta>0$,}
\\
>1 &\text{if $\beta<0$.}
\end{cases}
\ee
%
(We recall that $0<\gamma<|\beta|$.)
The phase $\theta$ winding is thus controled by the first term
of (\ref{chi(k)}) at $\beta>0$ and by the second term at $\beta<0$,
allowing us to write it as
\be\label{theta=-chi/2+}
\theta(k)=-\sgn \beta \frac{\chi(k)}2  +\Delta\theta(k)
,
\ee
where $\Delta\theta$ is an oscillatory periodic function of $\chi$.

\begin{figure}
\includegraphics[width=3.5in]{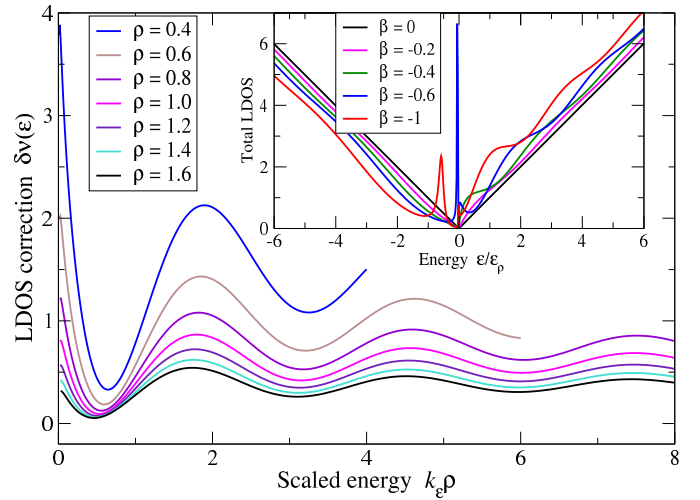} 
\vspace{-0.4cm}
\caption[]{
Standing wave oscillations in LDOS, Eq.(\ref{nu-general}), 
{\it vs.} energy scaled by 
the distance from impurity $\rho$.
Maxima (minima) occur at half-integer (integer) values of $k\rho/\pi$.
LDOS is shown for
an overcritical $\beta=0.6$ and several values of $\rho$, 
given in the units of $10^3r_0$.
{\it Inset:} Oscillations in LDOS, appearing for $|\beta|>\frac12$
on the top of the unperturbed density of states, which is subtracted
in the main figure ($\epsilon_\rho=\hbar v_F/\rho$).
}
\label{fig:LDOS}
\end{figure}

It is instructive to compare the behavior of $\Delta\theta(k)$ at 
strongly overcritical
and nearly critical $\beta$.
For large $|\beta|$ we have $\gamma \approx |\beta| $
and $|a|\approx e^{-2\pi\gamma}$, and thus the oscillatory
part $\Delta\theta$ is exponentially small:
\be
\Delta\theta(k) \approx \sgn\beta  e^{-2\pi\gamma} \sin(\arg  a -\chi(k))
.
\ee
%
In the opposite limit of nearly critical $\beta \approx 
\beta_c=\pm |m+\frac12|$,
we expand $|a|$ in $\gamma$, which is small in this case, 
to find
\be
|a|=1-\xi,\quad
\xi 
=\frac{2\pi\gamma}{1 - e^{-2\pi\beta_c}}+O(\gamma^2)
.
\ee
In this case $\theta$, as a function of $\chi$, is a staircase
\be
\theta = \frac{\zeta}{2} 
   - \arctan\lp \xi \tan\frac{\chi + \zeta}{2}\rp
,\quad
\zeta=\arg a 
,
\ee
with steps of height $\pi$, width $2\pi$, and 
corners rounded on the scale $\xi$.
The staircase slope $-\frac12\sgn\beta$ corresponds to the first term 
in Eq.(\ref{theta=-chi/2+}). 
The oscillatory part $\Delta\theta(k)$ manifests itself 
in the local density of states around the impurity 
(see peak in the $\epsilon<0$ region in Fig.\ref{fig:LDOS} inset).

To analyze the contribution of the phase $\theta(k)$ to the polarization
density, we suppress the periodic part $\Delta\theta$. Using
the expression (\ref{chi(k)}) we find $\theta(k) = - \gamma\sgn\beta \ln 2k r_0$.
Substituting it in the Friedel sum rule, Eq.(\ref{eq:Friedel_sum_rule}), 
we find
%
%
\begin{equation}
Q_{\rm pol}(\rho) = -N\frac{\theta(k\sim 1/\rho)}{\pi} =  
-\sgn\beta\frac{\gamma N}{\pi} \ln \frac{\rho}{2r_0}
. 
\end{equation}
%
From $n_{\rm pol} (\rho) = (2\pi \rho)^{-1} dQ_{\rm pol}/d\rho$,
we find the polarization density (\ref{eq:n_pol}).
%
%
%
When the parameter~$s$ is complex in more than one
channel, one has to consider a sum
%
\begin{equation}\label{eq:many_channels}
n_{\rm pol}(\rho) = -\frac{N\sgn\beta}{2\pi^2 \rho^2} 
\sum\limits_{|{m}+\frac12|<|\beta|}
\sqrt{\beta^2 - \left({m} + {\textstyle \frac12}\right)^2}
.
\end{equation}
%
For large~$\beta\gg1$, replacing the sum by an integral,
we recover the expression 
$n_{\rm pol}(\rho) = N \beta|\beta|/(4\pi\rho^2)$ 
found in \cite{Katsnelson06} by the Thomas--Fermi method. 


For supercritical $\beta>\frac12$, as discussed above, the scattering 
phase becomes sensitive to the physics at small distances,~$\rho \approx r_0$. 
This leads to pronounced interference of the incoming and outgoing waves, 
which is manifest in the local density of states (LDOS), 
%
\begin{equation}
\label{nu-general}
\nu (\epsilon, \rho) 
= \frac{N}{\pi \hbar v_F}\sum_m |\psi (k_\epsilon, \rho)|^2
,\quad
k_\epsilon = - \frac{\epsilon}{\hbar v_F}
,
\end{equation}
with an appropriate normalization
of the two-component wave function $\psi$, Eq.(\ref{ansatz}).
We evaluated the sum in (\ref{nu-general}) numerically, 
using the expressions (\ref{v-fall}) and (\ref{eq:w(z)_general}).
For $0 < |\beta| < \frac12$
LDOS does not deviate too much from $\nu_{\beta = 0}
\propto |\epsilon|$ (see Fig.\ref{fig:LDOS} inset).
For supercritical $\beta$, however, LDOS develops  pronounced 
oscillations
in both position $\rho$ and energy $k_\epsilon$. 
The crossover from a non-oscillatory to oscillatory behavior 
at $\beta\approx \frac12$ becomes sharp at $\rho\gg r_0$.

The standing waves in LDOS (\ref{nu-general}) at $\beta>\frac12$
are different from Friedel oscillations, since $k_F=0$ in our case
(Fermi level at the Dirac point).
As illustrated in Fig.\ref{fig:LDOS}, 
their spatial period scales inversely with energy,
so that the maxima occur at $k_\epsilon\rho=(n+\frac12)\pi$,
which is similar to the oscillations in LDOS studied 
in carbon nanotubes\,\cite{Ouyang}.
As in Ref.\cite{Ouyang}, the energy-dependent spatial period
can be used to obtain direct information about Fermi velocity $v_F$
in graphene.

The spatial structure predicted around supercritical 
Coulomb impurities, which extends up to the vacuum polarization cloud size 
$\rho_\star$, will be affected by finite temperature 
$T\gtrsim T_\ast=\hbar v_F/\rho_\star$,
and also by 
carrier doping away from neutrality by 
$\delta n\gtrsim\rho_\star^{-2}$
strong enough to induce screening at distances less than $\rho_\star$.

To summarize, we found that the excess charge $\beta-\frac12$
of supercritical impurities in graphene is fully screened
by the Dirac vacuum polarization. The large screening cloud size and the
standing wave oscillations predicted within it
can be directly probed by STM technique.
The sharp departure fom linear screening for supercritical impurities
represents an interesting example of nonlinear screening 
that can be realized in graphene. Our estimates for the critical charge, 
using $\kappa_{\rm RPA}\approx 5$,
yield an
experimentally convenient value $Z_c\sim 1$,
making experimental tests of these effects in graphene practical.




This work is supported by the DOE
(contract DEAC 02-98 CH 10886),
FOM (The Netherlands), NSF MRSEC (DMR 02132802) and
NSF-NIRT DMR-0304019.


\vspace{-0.5cm}

\begin{thebibliography}{99}

\bibitem{reviewGK}  A. K. Geim and K. S. Novoselov, Nat. Mater. \textbf{6},
183 (2007).

\bibitem{Gonzalez94} J. Gonz\'alez, F. Guinea, and M. A. H. Vozmediano, 
Nucl. Phys. B 424, 595 (1994); J. Low Temp. Phys. 99, 287 (1995)





\bibitem{nomura} K. Nomura and A. H. MacDonald, Phys. Rev. Lett.
{\bf 96}, 256602 (2006).

\bibitem{ando} T. Ando, J. Phys. Soc. Japan {\bf 75}, 074716 (2006).

\bibitem{dassarma} E. H. Hwang, S. Adam, and S. Das Sarma,
Phys. Rev. Lett. {\bf 98}, 186806 (2007).

\bibitem{Novoselov05}  K. S. Novoselov {\em et al.}
Nature
\textbf{438}, 197 (2005).

\bibitem{mele} D. P. DiVincenzo and E. J. Mele,
Phys. Rev. B {\bf 29}, 1685 (1984).


\bibitem{Katsnelson06} M. I. Katsnelson, Phys. Rev. B {\bf 74},
201401(R) (2006).

\bibitem{mirlin} P. M. Ostrovsky, I. V. Gornyi, and A. D. Mirlin,
Phys. Rev. B {\bf 74}, 235443 (2006).



\bibitem{Pomeranchuk}
I. Pomeranchuk and Y. Smorodinsky, J. Fiz. USSR 9, 97 (1945).

\bibitem{Zeldovich} 
Y. B. Zeldovich and V. S. Popov, 
Usp. Fiz. Nauk 105, 403 (1971); Eng. trans.:
Sov. Phys. Usp. 14, 673 (1972).

\bibitem{LL-3} 
L. D. Landau and E. M. Lifshitz, Quantum Mechanics,
Chap. XVII, \S 135 (3rd ed., Pergamon, London 1977). 






\bibitem{AbramowitzStegun} 
M. Abramowitz and I. A. Stegun, Handbook of Mathematical Functions, 
(Dover, New York, 1964). 

\bibitem{identity} 
\mbox{$z\,{{}_1F_{1}}' (\alpha, \gamma, z)=\alpha \left( {}_1F_{1} (\alpha + 1, \gamma, z) 
           - {}_1F_{1}(\alpha, \gamma, z)
          \right)$},
which follows from Eqs. (13.4.8), (13.4.4) of Ref.\cite{AbramowitzStegun}

\bibitem{asymptotic}
${}_1F_{1} (a, c; z) \approx  \frac{\Gamma (c)}{\Gamma (c - a)} (-z)^{-a}
 + \frac{\Gamma (c)}{\Gamma (a)} e^{z} z^{a - c} $
for large $|z|$ (see Eq.(13.5.1), Ref.\cite{AbramowitzStegun})

\bibitem{Friedel}
J. Friedel, Phil. Mag. 43, 153 (1952). 

\bibitem{Lin} 
D.-H. Lin, Phys. Rev. A 72, 012701 (2005); {\it ibid.} 73, 052113 (2006).

\bibitem{SSC} 
D. A. Abanin, P. A. Lee, and L. S. Levitov, 
{\it Solid State Comm.} {\bf 143}, 77 (2007)


\bibitem{Ouyang}
M. Ouyang, J.-L. Huang, and C. M. Lieber,
Phys. Rev. Lett. 88, 066804 (2002).



\end{thebibliography}
\end{document}